\documentstyle[preprint,aps]{revtex}
\tighten
\title{Improved variational principle for bounds on energy dissipation
    in turbulent shear flow}
\author{Rolf Nicodemus\footnote{Electronic address:
	    nicodemu@mailer.uni-marburg.de},
        Siegfried Grossmann\footnote{Electronic address:
	    grossmann\_s@stat.physik.uni-marburg.de},
	and Martin Holthaus\footnote{Electronic address:
	    holthaus@stat.physik.uni-marburg.de}}
\address{Fachbereich Physik der Philipps-Universit\"at, \\
    Renthof 6, D--35032 Marburg, Germany}
\date{August 30, 1996}
\narrowtext
\newenvironment{deflist}[1]%
{\begin{list}{}%
{\settowidth{\labelwidth}{\bf #1}%
\setlength{\leftmargin}{\labelwidth}%
\addtolength{\leftmargin}{\labelsep}%
}}%
{\end{list}}%
\begin{document}
\maketitle
\begin{abstract}
We extend the Doering-Constantin approach to upper bounds on energy
dissipation in turbulent flows by introducing a balance parameter into the
variational principle. This parameter governs the relative weight of
dif\-fer\-ent contributions to the dissipation rate. Its optimization leads to
improved bounds without entailing additional technical difficulties. For
plane Couette flow, the high-$Re$-bounds obtainable with one-dimensional
background flows are methodically lowered by a factor of $27/32$.

\vspace{2mm}


\end{abstract}

\vspace{2mm}


\section{Introduction}

While it is not feasible to obtain exact solutions to the equations of
motion for turbulent flows, it is possible to derive mathematically rigorous
upper bounds on certain quantities characterizing turbulent flow fields, such
as the rate of energy dissipation in shear flows or the rate of heat
transport in turbulent convection. A detailed theory of upper bounds has been
developed by Howard~\cite{Howard72} and Busse~\cite{Busse70,Busse78,Busse96}.
The bounds obtained by their method, although still considerably higher than
experimentally observed values, have until now resisted any attempt of
further improvement (see, e.g.,~\cite{KerswellSoward96}).

Recently a quite different approach has sparked renewed interest in the
theory of upper bounds. Instead of starting from the usual Reynolds
decomposition, Doering and
Constantin~\cite{DoeringConstantin92,DoeringConstantin94} utilized an idea
put forward already by Hopf~\cite{Hopf41} and decomposed a turbulent flow
field into a stationary ``background flow'' and a fluctuating component. The
background flow should be regarded as an arbitrary mathematical auxiliary
field that merely has to carry the boundary conditions of the physical flow
field, rather than as a time average of the actual flow. This method turns
out to be fairly
versatile~\cite{DoeringConstantin94,Marchioro94,GebhardtEtAl95,%
ConstantinDoering95,DoeringConstantin96}.
A given background flow immediately yields an upper bound, provided it
satisfies a certain spectral constraint. The best bounds obtainable in this
way can therefore be computed from a variational principle for the background
flow~\cite{DoeringConstantin94}. A formal connection between this approach
and the Howard-Busse theory has been elucidated by Kerswell~\cite{Kerswell96}.

The objective of the present paper is to point out that the variational
principle suggested by Doering and Constantin can still be improved, without
introducing additional technical complications. Although our arguments are of
a more general kind, we restrict the discussion to the concrete example of
energy dissipation in plane Couette flow. In this case the benefits of our
formulation of the variational principle are particularly obvious: the laminar
flow now becomes an admissible background flow for Reynolds numbers up to the
energy stability border, which is just what should be required on intuitive
grounds. This eliminates a shortcoming of the original Doering-Constantin
approach. For asymptotically high Reynolds numbers, and employing
one-dimensional background flows only, the improved principle yields upper
bounds on the rate of energy dissipation that are systematically by a factor
of $27/32$ lower than those that can be computed from the original principle.

\section{Formulation of the variational principle}

We consider an incompressible fluid confined between two infinitely extended
rigid plates. The lower plate at $z = 0$ is at rest, whereas the upper one at
$z = h$ moves with constant velocity $U$ in the positive $x$-direction. The
velocity field ${\bf u} \! \left( {\bf x},t \right)$ satisfies the
Navier-Stokes equations
\begin{equation}
  \partial_{t} {\bf u} + {\bf u} \cdot {\bf \nabla u} + {\bf \nabla} p
  = \nu \Delta {\bf u}
\label{ZWEI_NSEQ}
\end{equation}
with
\begin{equation}
  {\bf \nabla} \cdot {\bf u} = 0 \; ,
\label{ZWEI_DIV}
\end{equation}
where $p$ is the kinematic pressure and $\nu$ the kinematic viscosity. We
impose no-slip boundary conditions (b.c.) on ${\bf u}$ at $z = 0$ and
$z = h$,
\begin{equation}
  {\bf u} \! \left(x,y,0,t\right) = {\bf 0} \; , \;
  {\bf u} \! \left(x,y,h,t\right) = U {\bf \hat{x}}
\end{equation}
(${\bf \hat{x}}$ is the unit vector in $x$-direction), and periodic b.c.\ on
${\bf u}$ and $p$ in $x$- and $y$-direction; the periodicity lengths are
$L_{x}$ and $L_{y}$. The time-averaged energy dissipation rate per mass is
given by
\begin{equation}
  \varepsilon_{T} \equiv
  \left\langle\left\langle \nu \left|{\bf \nabla u}\right|^{2}
  \right\rangle\right\rangle_{T} =
  \frac{1}{T} \int_{0}^{T} \!\! dt \left[\frac{\nu}{\Omega}
  \int_{\Omega} \! d^{3}x \, u_{i|j} u_{i|j} \right] \; .
\end{equation}
We employ the notation
$\left\langle \cdot \right\rangle = \frac{1}{\Omega}
\int_{\Omega} \! d^{3}x \left(\cdot\right)$
for the volume average and
$\left\langle \cdot \right\rangle_{T}
= \frac{1}{T} \int_{0}^{T} \! dt \left(\cdot\right)$
for the time average, $\Omega = L_{x} L_{y} h$ denotes the periodicity
volume, and $u_{i|j}$ symbolizes $\partial_{j} u_{i} \! \left(x,y,z,t\right)$
with $i,j = x,y,z$; summation over repeated indices is implied. The goal
is to formulate a variational principle for rigorous upper bounds on the long
time limit of $\varepsilon_{T}$,
\begin{equation}
  \varepsilon \equiv \limsup_{T \rightarrow \infty} \varepsilon_{T} \; ,
\label{ZWEI_EPS}
\end{equation}
or on the non-dimensionalized quantity
\begin{equation}
  c_{\varepsilon} \! \left(Re\right) \equiv \frac{\varepsilon}{U^{3} h^{-1}}
  \; ,
\end{equation}
where $Re = U h / \nu$ is the Reynolds number.

The approach by Doering and
Constantin~\cite{DoeringConstantin92,DoeringConstantin94} rests on a
decomposition of the velocity field ${\bf u} \! \left({\bf x},t\right)$ into
a {\em background flow} and {\em fluctuations},
\begin{equation}
  {\bf u} \! \left({\bf x},t\right) = {\bf U} \! \left({\bf x}\right) +
  {\bf v} \! \left({\bf x},t\right) \; .
\end{equation}
The background flow ${\bf U}$ is a stationary and divergence-free vector
field satisfying the {\em physical b.c.}:
${\bf U} \! \left(x,y,0\right) = {\bf 0}$,
${\bf U} \! \left(x,y,h\right) = U {\bf \hat{x}}$,
and ${\bf U}$ is assumed to be periodic in $x$- and $y$-direction; otherwise it
is completely arbitrary. For the divergence-free fluctuations ${\bf v}$ we then
have {\em homogeneous b.c.} for all instants $t \geq 0$, i.e.,
${\bf v} \! \left(x,y,0,t\right) =
{\bf v} \! \left(x,y,h,t\right) = {\bf 0}$,
and ${\bf v}$ is periodic in $x$- and $y$-direction. With the usual
manipulations one arrives at the following equations for ${\bf u}$ and
${\bf v}$:
\begin{equation}
  \nu \left\langle \left|{\bf \nabla u} \right|^{2} \right\rangle =
  \nu \left\langle \left|{\bf \nabla U} \right|^{2} \right\rangle +
  2 \, \nu \left\langle U_{i|j} \, v_{i|j} \right\rangle +
  \nu \left\langle \left|{\bf \nabla v} \right|^{2} \right\rangle \; ,
\label{ZWEI_DISS}
\end{equation}
\begin{equation}
  \frac{d}{dt} \left\langle \frac{1}{2} \, {\bf v}^{2} \right\rangle +
  \left\langle {\bf v} \cdot \left({\bf U} \cdot {\bf \nabla U}\right)
  \right\rangle +
  \left\langle {\bf v} \cdot \left({\bf \nabla U}\right)_{\rm sym}
  \cdot {\bf v} \right\rangle = -
  \nu \left\langle U_{i|j} \, v_{i|j} \right\rangle -
  \nu \left\langle \left|{\bf \nabla v} \right|^{2} \right\rangle \; ,
\label{ZWEI_FLUCT}
\end{equation}
where
$ \left[\left({\bf \nabla U}\right)_{\rm sym}\right]_{i \, j} \equiv
\left(U_{j|i} + U_{i|j}\right) \! / 2$.
At this point, Doering and Constantin use eq.~(\ref{ZWEI_DISS}) to eliminate
the cross background-fluctuation term in eq.~(\ref{ZWEI_FLUCT}), namely, they
form
$\left[\mbox{\rm eq.~(\ref{ZWEI_DISS})}\right]
+ 2 \cdot \left[\mbox{\rm eq.~(\ref{ZWEI_FLUCT})}\right]$.
It is crucial to note that this implies putting a certain fixed weight on the
different contributions to the resulting expression for $\varepsilon_{T}$. In
order to avoid such a weighting we introduce a new degree of freedom and
consider
\begin{equation}
  \left[\mbox{\rm eq.~(\ref{ZWEI_DISS})}\right]
  + a \cdot \left[\mbox{\rm eq.~(\ref{ZWEI_FLUCT})}\right]
  \;\; , \;\; a > 1
\label{ZWEI_IDEA}
\end{equation}
with the {\em balance parameter} $a$. Apart from this modification, we now
follow closely the spirit of the original Doering-Constantin approach, which
can always be recovered by setting $a = 2$. Equation~(\ref{ZWEI_IDEA}) leads
to the {\em energy balance equation}
\begin{eqnarray}
  \varepsilon_{T} & = &
  \nu \left\langle \left|{\bf \nabla U} \right|^{2} \right\rangle -
  \frac{a}{T} \left\langle \frac{1}{2} \, {\bf v} \! \left(\cdot,T\right)^{2}
  \right\rangle +
  \frac{a}{T} \left\langle \frac{1}{2} \, {\bf v} \! \left(\cdot,0\right)^{2}
  \right\rangle
  \nonumber \\
  & - & a \, \left\langle \frac{1}{\Omega} \int_{\Omega} \! d^{3}x
  \left[ \frac{a-1}{a} \, \nu \left|{\bf \nabla v}\right|^{2} +
  {\bf v} \cdot \left({\bf \nabla U}\right)_{\rm sym} \cdot {\bf v} +
  {\bf f} \cdot {\bf v} \right] \right\rangle_{T}	
\end{eqnarray}
with
\begin{equation}
  {\bf f} \equiv {\bf U} \cdot {\bf \nabla U} -
  \frac{a-2}{a} \, \nu \Delta {\bf U} \; ,
\label{ZWEI_INHOM}
\end{equation}
which becomes the inequality (see~(\ref{ZWEI_EPS}))
\begin{eqnarray}
  \varepsilon & \leq &
  \nu \left\langle \left|{\bf \nabla U} \right|^{2} \right\rangle
  \nonumber \\
  & - & a \, \liminf_{T \rightarrow \infty }
  \left\langle \frac{1}{\Omega} \int_{\Omega} \! d^{3}x
  \left[ \frac{a-1}{a} \, \nu \left|{\bf \nabla v}\right|^{2} +
  {\bf v} \cdot \left({\bf \nabla U}\right)_{\rm sym} \cdot {\bf v} +
  {\bf f} \cdot {\bf v} \right] \right\rangle_{T} \; .
\label{ZWEI_EPSINEQ}
\end{eqnarray}
For the evaluation of~(\ref{ZWEI_EPSINEQ}) we have to distinguish two
different cases.

\vspace{3mm}

\noindent {\bf Case a) ${\bf f}$ is a gradient}

\vspace{3mm}

\noindent In this case the term linear in ${\bf v}$ on the rhs.\
of~(\ref{ZWEI_EPSINEQ}) vanishes and the resulting bound on the energy
dissipation rate is given by
\begin{equation}
  \varepsilon \leq \inf_{{\bf U}, \, a > 1}
  \left\{ \nu \left\langle \left|{\bf \nabla U} \right|^{2} \right\rangle
  \right\} \; ,
\label{ZWEI_EST1}
\end{equation}
{\em provided} ${\bf U}$ complies with the following constraints:
\begin{deflist}{iii)}
\item[i)] ${\bf U}$ is a divergence-free vector field satisfying the physical
b.c.\ and ${\bf f}$ is a gradient,
\item[ii)] the functional
\begin{equation}
  H_{{\bf U}, \, a} \! \left\{{\bf w}\right\} \equiv
  \frac{1}{\Omega} \int_{\Omega} \! d^{3}x
  \left[\frac{a-1}{a} \, \nu \left|{\bf \nabla w} \right|^{2} +
  {\bf w} \cdot \left({\bf \nabla U}\right)_{\rm sym} \cdot {\bf w} \right]
\label{ZWEI_FUNCTIONAL}
\end{equation}
is positive semi-definite,
$H_{{\bf U}, \, a} \! \left\{{\bf w}\right\} \geq 0$, for all stationary
divergence-free vector fields ${\bf w}$ satisfying the homogeneous b.c.
\end{deflist}

\noindent The condition {\bf ii} is equivalent to the statement that all
eigenvalues $\lambda$ of the hermitian eigenvalue problem
\begin{eqnarray}
  \lambda \, {\bf V} & = & - 2 \, \frac{a-1}{a} \, \nu \Delta {\bf V} +
  2 \left({\bf \nabla U}\right)_{\rm sym} \cdot {\bf V} + {\nabla} P \; ,
  \nonumber \\
  0 & = & {\bf \nabla} \cdot {\bf V} \;\; , \;\;
  \mbox{${\bf V}$ satisfies the homogeneous b.c.} \; ,
\label{ZWEI_EVAP}
\end{eqnarray}
are non-negative; $P$ is a Lagrange multiplier for the divergence condition.
Following Doering and Constantin~\cite{DoeringConstantin94}, who consider the
eigenvalue problem without an adjustable parameter ($a = 2$), we denote the
requirement that all eigenvalues of the eigenvalue problem~(\ref{ZWEI_EVAP})
be non-negative as
\begin{equation}
  \mbox{{\em spectral constraint:} all $\lambda \geq 0$} \; .
\label{ZWEI_SC1}
\end{equation}

\vspace{3mm}

\noindent {\bf Case b) ${\bf f}$ is no gradient}

\vspace{3mm}

\noindent In this case we bound the second term on the rhs.\ of
(\ref{ZWEI_EPSINEQ}) by
\begin{eqnarray}
  \lefteqn{\left\langle \frac{1}{\Omega} \int_{\Omega} \! d^{3}x
  \left[ \frac{a-1}{a} \, \nu \left|{\bf \nabla v}\right|^{2} +
  {\bf v} \cdot \left({\bf \nabla U}\right)_{\rm sym} \cdot {\bf v} +
  {\bf f} \cdot {\bf v} \right] \right\rangle_{T}} \makebox[2cm]{}
  \nonumber \\
  & \geq & \inf_{{\bf w}}
  \left\{\frac{1}{\Omega} \int_{\Omega} \! d^{3}x
  \left[ \frac{a-1}{a} \, \nu \left|{\bf \nabla w}\right|^{2} +
  {\bf w} \cdot \left({\bf \nabla U}\right)_{\rm sym} \cdot {\bf w} +
  {\bf f} \cdot {\bf w} \right]\right\} \; ,
\label{ZWEI_INEQ1}
\end{eqnarray}
where we seek the infimum in the space of all stationary, divergence-free
vector fields ${\bf w}$ satisfying the homogeneous b.c. If ${\bf V}$ is the
minimizing vector field, it solves the {\em Euler-Lagrange equations}
\begin{eqnarray}
  {\bf 0} & = & - 2 \, \frac{a-1}{a} \, \nu \Delta {\bf V} +
  2 \left({\bf \nabla U}\right)_{\rm sym} \cdot {\bf V} + {\nabla} P + {\bf f}
  \; , \nonumber \\
  0 & = & {\bf \nabla} \cdot {\bf V} \;\; , \;\;
  \mbox{${\bf V}$ satisfies the homogeneous b.c.} \; ,
\label{ZWEI_ELGL}
\end{eqnarray}
so that the inequality~(\ref{ZWEI_INEQ1}) becomes
\begin{equation}
  \left\langle \frac{1}{\Omega} \int_{\Omega} \! d^{3}x
  \left[ \frac{a-1}{a} \, \nu \left|{\bf \nabla v}\right|^{2} +
  {\bf v} \cdot \left({\bf \nabla U}\right)_{\rm sym} \cdot {\bf v} +
  {\bf f} \cdot {\bf v} \right] \right\rangle_{T} \geq
  \frac{1}{2 \, \Omega} \int_{\Omega} \! d^{3}x \; {\bf f} \cdot {\bf V} \; .
\end{equation}
The requirement
\begin{equation}
  H_{{\bf U}, \, a} \! \left\{{\bf w}\right\} > 0 \;\;
  \mbox{for all divergence-free ${\bf w} \neq {\bf 0}$ satisfying the
  homogeneous b.c.}
\label{ZWEI_REQU}
\end{equation}
guarantees the uniqueness of the solution ${\bf V}$ to the Euler-Lagrange
equations~(\ref{ZWEI_ELGL}) and ensures that this solution indeed minimizes
the rhs.\ of~(\ref{ZWEI_INEQ1}). The uniqueness can be shown as follows: if
there were any other solution ${\bf W} \neq {\bf V}$, then the difference
${\bf W} - {\bf V}$ would be a nonvanishing eigensolution to the eigenvalue
problem~(\ref{ZWEI_EVAP}) with eigenvalue $\lambda = 0$. This contradicts the
requirement~(\ref{ZWEI_REQU}), which implies that all eigenvalues
of~(\ref{ZWEI_EVAP}) are strictly positive. The minimizing character of
${\bf V}$ stems from the convexity of the expression in curly
brackets on the rhs.\ of~(\ref{ZWEI_INEQ1}).

Putting all things together, we see that in this case b) the resulting bound
on the energy dissipation rate is given by
\begin{equation}
  \varepsilon \leq \inf_{{\bf U}, \, a > 1}
  \left\{ \nu \left\langle \left|{\bf \nabla U} \right|^{2} \right\rangle -
  \frac{a}{2} \left\langle {\bf V} \cdot
  \left({\bf U} \cdot {\bf \nabla U}\right) \right\rangle +
  \frac{a-2}{2} \, \nu
  \left\langle {\bf V} \cdot \Delta {\bf U} \right\rangle \right\} \; ,
\label{ZWEI_EST2}
\end{equation}
{\em provided}
\begin{deflist}{iii)}
\item[i)] ${\bf U}$ is a divergence-free vector field satisfying the physical
b.c.\ and ${\bf f}$ is no gradient,
\item[ii)] ${\bf V}$ is a solution to the Euler-Lagrange
equations~(\ref{ZWEI_ELGL}),
\item[iii)] the functional~(\ref{ZWEI_FUNCTIONAL}) is strictly positive
definite, $H_{{\bf U}, \, a} \! \left\{{\bf w}\right\} > 0$ for all
divergence-free vector fields ${\bf w} \neq {\bf 0}$ satisfying the
homogeneous b.c. Equivalently, all eigenvalues $\lambda$ of the eigenvalue
problem~(\ref{ZWEI_EVAP}) must be positive, so that now we have the
\begin{equation}
  \mbox{{\em spectral constraint:} all $\lambda > 0$} \; .
\label{ZWEI_SC2}
\end{equation}
\end{deflist}

\noindent The main technical problem encountered in the original
Doering-Constantin approach is to verify their spectral constraint for a
given background flow, i.e., to show that all eigenvalues of the eigenvalue
problem~(\ref{ZWEI_EVAP}) with $a = 2$ are non-negative (case a) or positive
(case b). In this respect our formulation generates no additional
complications because our constraining eigenvalue problem is formally
identical with the original one, the only difference being that the kinematic
viscosity is rescaled by a factor $2 \left(a-1\right) \! / a$. Since
$ - \Delta$ is a positive definite, unbounded operator, the spectral constraint
enforces the positivity of this scaling factor, as anticipated in
eq.~(\ref{ZWEI_IDEA}) by the condition $a > 1$.

It should be noted that our formulation differs from the original one in two
major points. Firstly, if $\Delta {\bf U}$ is no gradient then the minimizing
field ${\bf V}$ must be determined by solving the Euler-Lagrange
equations~(\ref{ZWEI_ELGL}) even if ${\bf U} \cdot {\bf \nabla U}$ is a
gradient, since the contribution to ${\bf f}$ that is proportional to
$\Delta {\bf U}$ vanishes only if $a = 2$, see eq.~(\ref{ZWEI_INHOM}).
Secondly, our variational principle concerns not only the background flow
${\bf U}$ but also the balance parameter $a$. We will see that in the case of
one-dimensional background flows both issues can be dealt with easily; solving
the Euler-Lagrange equations and minimizing over $a$ will be independent from
the intricate problem of verifying the spectral constraint. Hence, the
additional freedom gained by the balance parameter $a$ entails no additional
difficulties, but will result in an improved bound on $\varepsilon$.

\section{Plane Couette flow with 1d background flows}

To become more specific we restrict the following discussion to
one-dimensional background flows, i.e., to flows which can be described
by a {\em profile function} ${\phi}$,
\begin{equation}
  {\bf U} = U \phi \! \left(\zeta\right) {\bf \hat{x}} \; ; \;\;
  \phi \! \left(0\right) = 0 \; , \; \phi \! \left(1\right) = 1
\label{DREI_PHIBC} \; ,
\end{equation}
depending on the dimensionless variable $\zeta \equiv z / h$. Clearly, such a
${\bf U}$ is a  divergence-free vector field satisfying the physical b.c.
Additionally we impose
$\phi \! \left(\zeta\right) = 1 -  \phi \! \left(1 - \zeta\right)$, so that
the profile is adapted to the symmetry of the physical problem.

\vspace{3mm}

\noindent {\bf a) Laminar profile, $\phi \! \left(\zeta\right) = \zeta$}

\vspace{3mm}

\noindent In this case ${\bf U}$ is the laminar solution to the equations of
motion~(\ref{ZWEI_NSEQ}) and~(\ref{ZWEI_DIV}), and we have
${\bf f} = {\bf 0}$. Hence, case a) of our variational principle applies.
Provided all eigenvalues of the eigenvalue problem~(\ref{ZWEI_EVAP}) are
non-negative, the estimate~(\ref{ZWEI_EST1}) yields
\begin{equation}
  \varepsilon = \nu \, \frac{U^{2}}{h^{2}} \;\;\;\; \mbox{or} \;\;\;\;
  c_{\varepsilon} = Re^{-1} \; ,
\label{DREI_EPSLAM}
\end{equation}
respectively. We have strict equalities here since upper and lower bounds on
the long time limit of $\varepsilon_{T}$ coincide, as long as the laminar
profile fulfills the spectral constraint~\cite{DoeringConstantin94}. We are
thus led to the following question: what is the maximal Reynolds number~$Re$
(or inverse kinematic viscosity $\nu^{-1}$) up to which the laminar profile
is admitted as a valid testprofile for the variational principle?

Since the viscosity enters into the eigenvalue problem~(\ref{ZWEI_EVAP})
only in rescaled form, we can tune the Reynolds number by suitably
adjusting the value of $a$. In order to find the maximal ``critical''
Reynolds number for the laminar profile we have to set $a = \infty$, so that
the scaling factor $2 \left(a-1\right) \! / a$ of the viscosity takes on the
highest value possible, namely $2$. The eigenvalue problem then becomes
\begin{eqnarray}
  \lambda \, {\bf V} & = & - 2 \, \nu \Delta {\bf V} + \frac{U}{h}
  \left(
  \begin{array}{ccc} 0 & 0 & 1 \\ 0 & 0 & 0 \\ 1 & 0 & 0 \end{array}\right)
  \cdot {\bf V} + {\nabla} P \; , \nonumber \\
  0 & = & {\bf \nabla} \cdot {\bf V} \;\; , \;\;
  \mbox{${\bf V}$ satisfies the homogeneous b.c.}
\end{eqnarray}
This is exactly the eigenvalue problem appearing in energy stability theory
for the plane Couette flow~\cite{Joseph76,DrazinReid81}. Hence, the laminar
profile is an admissible testprofile up to the Reynolds number
$Re_{\, \rm ES}$ characterizing the energy stability
border~\cite{Joseph76,DrazinReid81},
\begin{equation}
  Re_{\, \rm ES} \simeq 2 \, \sqrt{1707.77} \simeq 82.65 \; ,
\label{DREI_REES}
\end{equation}
and consequently~(\ref{DREI_EPSLAM}) is valid for all
$Re \leq Re_{\, \rm ES}$.

In the original Doering-Constantin approach ($a = 2$) one had to discard the
laminar profile already for Reynolds numbers $Re > Re_{\, \rm ES} / 2$, which
necessarily led to non-optimal upper bounds on $c_{\varepsilon}$ for Reynolds
numbers $Re_{\, \rm ES} / 2 < Re \leq Re_{\, \rm ES}$. The introduction of
the balance parameter cures this obvious shortcoming and guarantees that
$c_{\varepsilon}$ as obtained from the variational principle has the known
$1 / Re$-behaviour up to $Re = Re_{\, \rm ES}$.

\vspace{3mm}

\noindent {\bf b) Non-laminar profile,
$\phi \! \left(\zeta\right) \neq \zeta$}

\vspace{3mm}

\noindent In this case we have
\begin{equation}
  {\bf f} = - \frac{a-2}{a} \, \nu \, \frac{U}{h^{2}} \,
  \phi^{\prime \prime} \! \left(\zeta\right) {\bf \hat{x}} \;\;\;\;
  \mbox{with} \;\;\;\; \phi^{\prime \prime} \! \left(\zeta\right) \neq 0 \; .
\end{equation}
The boundary conditions~(\ref{DREI_PHIBC}) together with the symmetry
condition ensure that ${\bf f}$ cannot be a gradient if $a \neq 2$,
so that now we have to resort to case b) of our variational principle.
Because of the special form that the inhomogeneous term ${\bf f}$ acquires
for non-laminar profiles, one can find an analytical solution to the
Euler-Lagrange equations~(\ref{ZWEI_ELGL}):
\begin{eqnarray}
  {\bf V} & = & \frac{1}{2} \, \frac{a-2}{a-1} \, U
  \left[\zeta - \phi \! \left(\zeta\right) \right] {\bf \hat{x}} \; ,
\label{DREI_SOLV} \\
  P & = & P_{0} + \frac{1}{2} \, \frac{a-2}{a-1} \, U^{2}
  \left[\frac{1}{2} \, \phi \! \left(\zeta\right)^{2} -
  \int_{0}^{\zeta} \!\! d\xi \, \xi \, \phi^{\prime} \! \left(\xi\right)
  \right] \; .
\end{eqnarray}
Note that ${\bf V}$ vanishes if $a = 2$, so that we do not need to distinguish
between the cases $a \neq 2$ and $a = 2$. Moreover, the rhs.\
of~(\ref{DREI_SOLV}) is proportional to the difference between
$\phi \! \left(\zeta\right)$ and the the laminar profile $\zeta$, which
leads us back to the laminar case a) in the limit
$\phi \! \left(\zeta\right) \rightarrow \zeta$. Provided all eigenvalues
of~(\ref{ZWEI_EVAP}) are positive, the estimate~(\ref{ZWEI_EST2}) yields
\begin{equation}
  \varepsilon \leq \inf_{\phi, \, a > 1} \left\{
  \left[1 + \frac{a^{2}}{4 \left(a - 1 \right)} \,
  D \! \left\{\phi\right\} \right]
  \nu \, \frac{U^{2}}{h^{2}} \right\} \; ;
\label{DREI_EPSNONLAM}
\end{equation}
the inequality for $c_{\varepsilon}$ reads
\begin{equation}
  c_{\varepsilon} \leq \inf_{\phi, \, a > 1} \left\{
  \left[1 + \frac{a^{2}}{4 \left(a - 1 \right)} \,
  D \! \left\{\phi\right\} \right] Re^{-1} \right\} \; .
\label{DREI_CEPSNONLAM}
\end{equation}
Here we have employed the abbreviation $D \! \left\{\phi\right\}$ for the
functional
\begin{equation}
  D \! \left\{\phi\right\} \equiv \int_{0}^{1} \!\! d\zeta
  \left[\phi^{\prime} \! \left(\zeta\right)\right]^{2} - 1 \; .
\label{DREI_D}
\end{equation}
This functional is strictly positive for
$\phi \! \left(\zeta\right) \neq \zeta$, which means that for each non-laminar
profile and each $a > 1$ the factor
$1 + \frac{a^{2}}{4 \left(a -1 \right)} \, D \! \left\{\phi\right\}$
appearing in~(\ref{DREI_EPSNONLAM}) and~(\ref{DREI_CEPSNONLAM}) exceeds~one.
Therefore, each non-laminar profile produces a bound that is strictly higher
than the laminar bound~(\ref{DREI_EPSLAM}).

In a manner analogous to the procedure for the laminar profile one has to
investigate on the basis of the eigenvalue problem~(\ref{ZWEI_EVAP}) up to
which Reynolds number a given profile $\phi$ is admissible as a testprofile.
Because of the positive definiteness of $- \Delta$, the spectrum of the
hermitian operator defined by the rhs.\ of~(\ref{ZWEI_EVAP}) is lowered
when the rescaled kinematic viscosity is decreased. Thus, for a given
$\phi$ we define the {\em critical Reynolds number}
$R_{c} \! \left\{\phi\right\}$
as that Reynolds number where the lowest eigenvalue $\lambda$
of~(\ref{ZWEI_EVAP}) with fixed balance parameter $a = \infty$
(i.e., $2 \left(a-1\right) \! / a = 2$ is maximal)
passes through zero. For $Re \geq R_{c} \! \left\{\phi\right\}$ one has to
discard $\phi$ as a testprofile for the variation. In addition, if $a > 1$
is finite, one finds the constraint
\begin{equation}
  Re < \frac{a-1}{a} \, R_{c} \! \left\{\phi\right\}	\; .
\label{DREI_CONSTR}
\end{equation}

The factor $a^{2}/ \! \left(a-1\right)$ in~(\ref{DREI_EPSNONLAM})
and~(\ref{DREI_CEPSNONLAM}) has a local minimum for $a = 2$ and increases
monotonically with $a$ for $a > 2$. A given $\phi$ thus produces the following
upper bound on $c_{\varepsilon}$:
\begin{equation}
  c_{\varepsilon} \leq \left[1 +
  \frac{a_{{\rm min}}^{2}}{4 \left(a_{{\rm min}} -1\right)} \,
  D \! \left\{\phi\right\}\right] Re^{-1} \;\;\;\; \mbox{for} \;\;\;\;
  0 \leq Re < R_{c} \! \left\{\phi\right\}
\end{equation}
with
\begin{equation}
  a_{{\rm min}} = \left\{
  \begin{array}{ll}
    2 & \;\; \mbox{for} \;\;\;\;
    0 \leq Re < \frac{1}{2} R_{c} \! \left\{\phi\right\} \\
    \frac{R_{c} \! \left\{\phi\right\}}{R_{c} \! \left\{\phi\right\} - Re} &
    \;\; \mbox{for} \;\;\;\;
    \frac{1}{2} R_{c} \! \left\{\phi\right\}
    \leq Re < R_{c} \! \left\{\phi\right\}
  \end{array} \right. \; ;
\label{DREI_AMIN}
\end{equation}
hence
\begin{equation}
  c_{\varepsilon} \leq \left\{
  \begin{array}{ll}
    \left[1 + D \! \left\{\phi\right\}\right] Re^{-1} & \;\; \mbox{for}
    \;\;\;\; 0 \leq Re < \frac{1}{2} R_{c} \! \left\{\phi\right\} \\
    \left[1 + \frac{D \! \left\{\phi\right\}
    R_{c} \! \left\{\phi\right\}^{2}}
    {4 \left(R_{c} \! \left\{\phi\right\} - Re\right) Re}\right] Re^{-1}
    & \;\; \mbox{for} \;\;\;\;
    \frac{1}{2} R_{c} \! \left\{\phi\right\}
    \leq Re < R_{c} \! \left\{\phi\right\}
  \end{array} \right. \; .
\label{DREI_CBOUND}
\end{equation}
In this way we have accomplished the optimization of the balance parameter and
are left with the task of varying the profile function. But it is possible
to deduce some general statements even {\em without} solving the variational
principle for $\phi$. To this end, we denote the expression on the rhs.\
of~(\ref{DREI_CBOUND}) as $\bar{c}_{\varepsilon} \! \left(Re\right)$.
This is a continuous function of the Reynolds number, and even continuously
differentiable at $Re = R_{c} \! \left\{\phi\right\} \! / 2$. For every given
profile, $\bar{c}_{\varepsilon} \! \left(Re\right)$ has exactly one local
minimum in the whole interval $0 \leq Re < R_{c} \! \left\{\phi\right\}$;
this minimum appears in the upper half interval
$R_{c} \! \left\{\phi\right\} \! / 2 \leq Re < R_{c} \! \left\{\phi\right\}$.
Because the variational principle tests {\em all} profile functions $\phi$
satisfying the required conditions, the minimum point determined by a
particular $\phi$ is the only point that this profile could possibly
contribute to the resulting upper bound on
$c_{\varepsilon} \! \left(Re\right)$. Thus, each $\phi$ leads to a point in
the ($Re, c_{\varepsilon}$)-plane,
\begin{equation}
  \phi \;\; \longrightarrow \;\;
  \left(R_{{\rm min}} \! \left\{\phi\right\},
  \bar{c}_{\varepsilon} \! \left( R_{{\rm min}} \! \left\{\phi\right\} \right)
  \right) \;.
\label{DREI_POINT}
\end{equation}
The Reynolds number $R_{{\rm min}} \! \left\{\phi\right\}$ of the minimum
point can be expressed as
\begin{equation}
  R_{{\rm min}} \! \left\{\phi\right\} = x_{0} \! \left\{\phi\right\}
  R_{c} \! \left\{\phi\right\} \; ,
\end{equation}
where $x_{0} \! \left\{\phi\right\}$ is the unique (real) zero
$x_{0}$ with $1/2 \leq x_{0} < 1$ of the cubic polynomial
\begin{equation}
  x^{3} - 2 \, x^{2} + \left(1 - \frac{3}{4} \,
  D \! \left\{\phi\right\}\right) x +
  \frac{1}{2} \, D \! \left\{\phi\right\} = 0 \; .
\label{DREI_POLY}
\end{equation}
This follows directly by minimizing
$\bar{c}_{\varepsilon} \! \left(Re\right)$ with respect to $Re$. Although an
analytical expression for the desired zero of~(\ref{DREI_POLY}) is available
for all $D \! \left\{\phi\right\} > 0$, we will discuss only two limiting
cases. In the {\em laminar limit}
$\phi \! \left(\zeta\right) \rightarrow \zeta$ the functional
$D \! \left\{\phi\right\}$ vanishes and the zero
$x_{0} \! \left\{\phi\right\}$ tends to~$1$. Inserting
$R_{{\rm min}} \! \left\{\phi\right\}$ into~(\ref{DREI_AMIN}) shows that
$a_{{\rm min}}$ tends to infinity in this limit,
\begin{equation}
  \lim_{Re \, \searrow \, Re_{\, \rm ES}} a_{{\rm min}} = \infty \; .
\end{equation}
Therefore, the optimal bounds on $c_{\varepsilon}$ resulting from the
variational principle will be continuous at the energy stability border. On
the other hand, for sufficiently high Reynolds numbers the spectral
constraint~(\ref{ZWEI_SC2}) singles out those profile functions as admissible
for the variational principle that have large slopes within thin boundary
layers and remain almost constant in the
interior~\cite{DoeringConstantin94}. Hence the functional
$D \! \left\{\phi\right\}$ tends to infinity in the {\em asymptotic limit}
$Re \rightarrow \infty$, which implies that the zero
$x_{0} \! \left\{\phi\right\}$ tends to~$2/3$ and $a_{{\rm min}}$ approaches
its asymptotic value~$3$,
\begin{equation}
  \lim_{Re \rightarrow \infty} a_{{\rm min}} = 3 \; .
\end{equation}
The optimal value of the balance parameter $a$, considered as a function of
the Reynolds number, is monotonically decreasing from infinity at the
energy stability border $Re_{\, \rm ES}$ to its asymptotic limit
$a_{\infty} = 3$ and never reaches the value $a = 2$ that has tacitly been
taken for granted before.

In the original approach~\cite{DoeringConstantin94}
a given profile $\phi$ yields an upper bound on $c_{\varepsilon}$ in a similar
way, but the range of Reynolds numbers within which this profile is admissible
is only half as large as in our case. Since, as remarked before, the factor
$a^{2}/ \! \left(a-1\right)$ appearing in~(\ref{DREI_CEPSNONLAM}) has a
minimum for $a = 2$, in the common  interval this bound agrees with ours:
\begin{equation}
  c_{\varepsilon} \leq \bar{c}_{\varepsilon} \! \left(Re\right) \;\;\;\;
  \mbox{for} \;\;\;\; 0 \leq Re \leq
  \frac{1}{2} \, R_{c} \! \left\{\phi\right\} \; .
\end{equation}
Here the minimizing point is
$\left(R_{c} \! \left\{\phi\right\} \! / 2 , \;
\bar{c}_{\varepsilon} \! \left(R_{c} \! \left\{\phi\right\} \! / 2
\right)\right)$,
which has to be compared with~(\ref{DREI_POINT}). Therefore, in our
formulation a given profile function $\phi$ leads to a {\em lowered} bound
on $c_{\varepsilon}$ at a {\em higher} Reynolds number $Re$. The improvement
due to the balance parameter $a$ relies on the fact that each $\phi$ becomes
admissible within an enlarged $Re$-interval. Note that in both cases exactly
the same eigenvalue problem must be solved to check the spectral constraint.

Under the assumption that in both formulations the upper bounds on
$c_{\varepsilon}$ are approaching constant values in the asymptotic limit
$Re \rightarrow \infty$, dubbed $\bar{c}_{\varepsilon, \, \infty}$ and
$\bar{c}_{\varepsilon, \, \infty}^{\, {\rm DC}}$, respectively, we can
calculate the relative factor
\begin{equation}
  g_{\infty} \equiv \frac{\bar{c}_{\varepsilon, \, \infty}}
  {\bar{c}_{\varepsilon, \, \infty}^{\, {\rm DC}}} =
  \lim_{Re \rightarrow \infty}
  \frac{\bar{c}_{\varepsilon} \! \left(x_{0} \! \left\{\phi\right\}
  R_{c} \! \left\{\phi\right\}\right)}
  {\bar{c}_{\varepsilon} \! \left(
  R_{c} \! \left\{\phi\right\} \! / 2 \right)} = \frac{27}{32} \; .
\end{equation}
Considering the same class of one-dimensional background flows, our
formulation yields asymptotic upper bounds on the rate of energy dissipation
that are systematically lowered by a factor of $27/32$ compared to those
that can be computed from the original Doering-Constantin approach.

\section{Illustration}

In this section we wish to illustrate the general statements derived in the
previous section. Our aim is not to calculate the best bounds that the
improved variational principle has to offer, but rather to elucidate the new
aspects in a simple way. Instead of verifying the spectral constraint for a
given $\phi$ on the basis  of the eigenvalue problem~(\ref{ZWEI_EVAP}), one
can check the positivity of the functional
$H_{{\bf U}, \, a} \! \left\{{\bf w}\right\}$,
\begin{equation}
  H_{{\bf U}, \, a} \! \left\{{\bf w}\right\} =
  \frac{1}{\Omega} \int_{\Omega} \! d^{3}x
  \left[\frac{a-1}{a} \, \nu \left|{\bf \nabla w} \right|^{2} +
  \frac{U}{h} \, \phi^{\prime} \, w_{x} w_{z} \right] \; ,
\end{equation}
by means of the inequality
\begin{equation}
  \left| \frac{1}{\Omega} \int_{\Omega} \! d^{3}x \,
  \frac{U}{h} \, \phi^{\prime} \, w_{x} w_{z} \right| \leq
  \frac{\left\langle \left|{\bf \nabla w} \right|^{2} \right\rangle}
  {2 \sqrt{2}}
  \int_{0}^{h/2} \!\! dz \, z \, \frac{U}{h} \left| \phi^{\prime} \right| \; .
\end{equation}
This estimate can be shown by repeatedly using Schwarz' inequality and
utilizing the symmetry property of the profile
function~\cite{GebhardtEtAl95} (see also
refs.~\cite{DoeringConstantin92,DoeringConstantin94}). Thus,
\begin{equation}
  H_{{\bf U}, \, a} \! \left\{{\bf w}\right\} \geq
  \left[\frac{a-1}{a} \, \nu - \frac{U h}{2 \sqrt{2}} \,
  \int_{0}^{1/2} \!\! d \zeta \, \zeta \left| \phi^{\prime} \right| \right]
  \left\langle \left|{\bf \nabla w} \right|^{2} \right\rangle \; .
\end{equation}
If the profile function $\phi \! \left(\zeta\right)$ satisfies the condition
\begin{equation}
  \int_{0}^{1/2} \!\! d \zeta \, \zeta \left|
  \phi^{\prime} \! \left(\zeta\right)\right| \leq 2 \sqrt{2} \,
  \frac{a-1}{a} \, Re^{-1} \;\;\;\; \left(\;\; \mbox{or} \;\;\;\;
  \int_{0}^{1/2} \!\! d \zeta \, \zeta \left|
  \phi^{\prime} \! \left(\zeta\right)\right| < 2 \sqrt{2} \,
  \frac{a-1}{a} \, Re^{-1} \right) \; ,
\label{VIER_CONSTR}
\end{equation}
then the non-negativity (or the positivity) of the functional
$H_{{\bf U}, \, a} \! \left\{{\bf w}\right\}$ is guaranteed. It should be
realized that~(\ref{VIER_CONSTR}) is a sufficient but not a necessary
condition; it is more restrictive than the spectral constraint.
In addition to the b.c.~(\ref{DREI_PHIBC}) and the symmetry requirement
we assume $\phi^{\prime} \! \left(\zeta\right) \geq 0$, so that the modulus
signs $\left| \right|$ in~(\ref{VIER_CONSTR}) can be skipped.

After the replacement of the variational principle's spectral constraint
(condition {\bf ii} in case a) and condition {\bf iii} in case b),
respectively) by the {\em sharpened profile constraint}~(\ref{VIER_CONSTR}),
a certain $\phi$ can, in general, no longer be admitted as a testprofile for
Reynolds numbers up to $R_{c} \! \left\{\phi\right\}$ as defined in the
previous section. Rather,~(\ref{VIER_CONSTR}) leads to  the border
\begin{equation}
  R_{c}^{\rm S} \! \left\{\phi\right\} \equiv \frac{2 \sqrt{2}}
 {\int_{0}^{1/2} \!\! d \zeta \, \zeta \,
 \phi^{\prime} \! \left(\zeta\right)} \; .
\end{equation}
For example, the laminar profile $\phi \! \left(\zeta\right) = \zeta$ yields
\begin{equation}
  R_{c}^{\rm S} \! \left\{\zeta\right\} = 16 \sqrt{2} \simeq 22.63,
\end{equation}
which has to be contrasted to~(\ref{DREI_REES}), i.e.,
$R_{c} \! \left\{\zeta\right\} = Re_{\, \rm ES} \simeq 82.65$. Thus
$R_{c}^{\rm S} \! \left\{\zeta\right\}$ is by about a factor~$4$ smaller
than $R_{c} \! \left\{\zeta\right\}$. Nevertheless, it is instructive to
consider the variational principle with the stronger
constraint~(\ref{VIER_CONSTR}) since then even the variation of the profile
$\phi$ can be done analytically. For finite $a > 1$ we now have the
constraint
\begin{equation}
  Re < \frac{a-1}{a} \, R_{c}^{\rm S} \! \left\{\phi\right\}
\end{equation}
instead of~(\ref{DREI_CONSTR}), which is exactly~(\ref{VIER_CONSTR}).

Without going into the cumbersome technical details (see also
ref.~\cite{GebhardtEtAl95} for comparison) we summarize our results.
We have to distinguish three $Re$-ranges,
\begin{equation}
  \left.
  \begin{array}{lcccccl}
    \left({\rm I}\right) & 0  & \leq & Re & \leq & 16 \sqrt{2} & , \\
    \left({\rm II}\right) & 16 \sqrt{2} & < & Re & \leq & 20 \sqrt{2} & , \\
    \left({\rm III}\right) & 20 \sqrt{2} & < & Re & < & \infty & .
  \end{array}
  \right.
\end{equation}
In these ranges the minimizing profile function behaves as sketched in
Fig.~1: in~(I) $\phi$ is equal to the laminar profile,
$\phi \! \left(\zeta\right) = \zeta$. In~(II) the profile develops parabolic
boundary layers which reach the middle of the domain~$\Omega$ (i.e.,
$\zeta = 1/2$), while the slope at $\zeta = 1/2$ decreases from~one to~zero
with increasing $Re$. When $Re$ is increased beyond $20\sqrt{2}$, the
thickness $\delta$ of these parabolic boundary layers is getting smaller and
smaller; in the interior $\phi$ remains constant,
$\phi \! \left(\zeta\right) = 1/2$ for $\delta \leq \zeta \leq 1 - \delta$.
Asymptotically, $\delta$ vanishes as $\sim 1/Re$. The analytic expression for
$\delta$ in~(III) reads
\begin{equation}
  \delta = \frac{4 \sqrt{2}}{Re} \,
  \frac{3 \sqrt{Re - 2 \sqrt{2}} + \sqrt{Re - 18 \sqrt{2}}}
  {\sqrt{Re - 2 \sqrt{2}} + \sqrt{Re - 18 \sqrt{2}}} \; \; \;
  \stackrel{Re \rightarrow \infty}{\longrightarrow} \; \; \;
  \frac{8 \sqrt{2}}{Re} \; .
\end{equation}
Correspondingly, the functional $D \! \left\{\phi\right\}$ defined by
eq.~(\ref{DREI_D}) diverges for $Re \rightarrow \infty$, as anticipated
in section~3.

The minimizing balance parameter $a$ turns out to be
\begin{equation}
  a = \left\{
  \begin{array}{lc}
    \infty \;\;\; \left(\mbox{i.e., $2 \, \frac{a-1}{a} = 2$}\right) &
    \;\;\;\; \left({\rm I}\right) \\
    2 + \frac{16 \sqrt{2}}{Re - 16 \sqrt{2}} &
    \;\;\;\; \left({\rm II}\right) \\
    \frac{3}{2}
    \left(1 + \sqrt{\frac{Re - 2 \sqrt{2}}{Re - 18 \sqrt{2}}} \, \right)
    & \;\;\;\; \left({\rm III}\right)
  \end{array}
  \right. \; .
\end{equation}
For $Re > 16 \sqrt{2}$ this parameter decreases monotonically from $\infty$
to its asymptotic value $a_{\infty} = 3$, as depicted in Fig.~2.

Finally, the resulting bounds on $c_{\varepsilon}$ are given by
\begin{equation}
  \frac{1}{Re} \leq c_{\varepsilon} \! \left(Re\right) \leq
  \left\{
  \begin{array}{lc}
    \frac{1}{Re} & \;\;\;\; \left({\rm I}\right) \\
    \left(1 - \frac{12 \sqrt{2}}{Re}\right) \frac{4}{Re} &
    \;\;\;\; \left({\rm II}\right) \\
    \frac{3}{16 \sqrt{2}}
    \frac{\sqrt{Re - 2 \sqrt{2}} + \sqrt{Re - 18 \sqrt{2}}}
    {3 \sqrt{Re - 2 \sqrt{2}} + \sqrt{Re - 18 \sqrt{2}}} +
    \frac{1}
    {\left(3 \sqrt{Re - 2 \sqrt{2}} + \sqrt{Re - 18 \sqrt{2}}\right)^{2}} &
    \;\;\;\; \left({\rm III}\right) \\
  \end{array}
  \right. \; .
\label{VIER_CBOUND}
\end{equation}
Note that the upper bound is continuous but not continuously differentiable
at $Re = 16 \sqrt{2}$, see Fig.~3. The limit $Re \rightarrow \infty$ yields
\begin{equation}
  c_{\varepsilon} \! \left(\infty\right) \leq
  \lim_{Re \rightarrow \infty} \bar{c}_{\varepsilon} \! \left(Re\right) =
  \frac{3}{32 \sqrt{2}} \; .
\end{equation}
As before, we denote the upper bound on $c_{\varepsilon}$ by
$\bar{c}_{\varepsilon} \! \left(Re\right)$.

These results have to be compared with those derived in
ref.~\cite{GebhardtEtAl95}, where the sharpened profile
constraint~(\ref{VIER_CONSTR}) was used to calculate bounds on
$c_{\varepsilon}$ analytically within the framework established by Doering
and Constantin ($a = 2$ fixed). The analogous three $Re$-ranges found there are
\begin{equation}
  \left.
  \begin{array}{lcccccl}
    \left({\rm I^{\prime}}\right) & 0  & \leq & Re & \leq & 8 \sqrt{2} & ,\\
    \left({\rm II^{\prime}}\right) & 8 \sqrt{2} & < & Re & \leq & 12 \sqrt{2}
    & , \\
    \left({\rm III^{\prime}}\right) & 12 \sqrt{2} & < & Re & < & \infty & ;
  \end{array}
  \right.
\end{equation}
in particular, the first range (${\rm I}^{\prime}$) is only half as wide as
in our case. The corresponding bounds on $c_{\varepsilon}$ are given by
\begin{equation}
  \frac{1}{Re} \leq c_{\varepsilon} \! \left(Re\right) \leq
  \left\{
  \begin{array}{lc}
    \frac{1}{Re} & \;\;\;\; \left({\rm I^{\prime}}\right) \\
    \left(1 - \frac{12 \sqrt{2}}{Re} + \frac{96}{Re^{2}}\right) \frac{4}{Re}
    & \;\;\;\; \left({\rm II^{\prime}}\right) \\
    \frac{1}{9 \sqrt{2}} & \;\;\;\; \left({\rm III^{\prime}}\right) \\
  \end{array}
  \right. \; .
\label{VIER_CBOUNDDC}
\end{equation}
The improvement due to the parameter $a$ is measured by the ratio of the
rhs.\ of~(\ref{VIER_CBOUND}) and the rhs.\ of~(\ref{VIER_CBOUNDDC}),
\begin{equation}
  \widetilde{g} \! \left(Re\right) \equiv
  \frac{\bar{c}_{\varepsilon} \! \left(Re\right)}
  {\bar{c}_{\varepsilon}^{\, {\rm DC}} \! \left(Re\right)} \; .
\end{equation}
This ratio has a minimum at $Re = 16 \sqrt{2}$,
\begin{equation}
  \widetilde{g} \! \left(16 \sqrt{2}\right) =
  \frac{9 \sqrt{2}}{16 \sqrt{2}} = \frac{9}{16} \simeq 0.56 \; ,
\end{equation}
and increases monotonically with increasing $Re$ to its asymptotic value
$27/32$,
\begin{equation}
  \lim_{Re \rightarrow \infty} \widetilde{g} \! \left(Re\right) =
  \frac{3 \cdot 9 \sqrt{2}}{32 \sqrt{2}} = \frac{27}{32} \simeq 0.84 \; .
\end{equation}
A graphical comparison of the bounds~(\ref{VIER_CBOUND}) and
(\ref{VIER_CBOUNDDC}) is shown in Fig.~3.

\section{Conclusion: background flow and balance parameter belong together}

In this paper we have been concerned with an improved {\em formulation} of
the Doering-Constantin variational principle for bounds on turbulent energy
dissipation, rather than with its {\em solution}. The necessity of devoting
the utmost care to an optimal formulation is obvious: what has been lost by a
non-optimal formulation of a variational principle can not be regained by
even the most sophisticated techniques for solving it.

The feasibility of improving the background flow method rests on two key
observations. The first of these is that if one refrains from what appears to
be straightforward, namely, if one does {\em not} eliminate the unwanted
cross background-fluctuation term from eq.~(\ref{ZWEI_FLUCT}) but rather
keeps this term with a certain weight quantified by the balance parameter
$a$, one gains a new freedom that can be exploited to improve the bounds. The
second observation is that the required solution to the Euler-Lagrange
equations~(\ref{ZWEI_ELGL}) can easily be written down analytically if one
restricts oneself to background flows which only have a height-dependent
profile but no spanwise structure. Then the technical difficulties
encountered in our formulation are the same as those of the original approach
--- in both cases one has to solve the same eigenvalue problem in order to
check whether a given profile is an admissible candidate for the computation
of the bound, albeit in our case the kinematic viscosity is rescaled ---, but
the improved formulation leads to systematically lowered bounds.

Finally, we wish to stress that the usefulness of introducing the balance
parameter is by no means restricted to the case of plane Couette flow. It
should be obvious by now that the arguments employed in sections~2 and~3 can
easily be adapted to other problems. When computing upper bounds on {\em any}
property of a turbulent flow that is amenable to the background flow method,
background flow and balance parameter should henceforth be regarded as
belonging together.

\vspace{3mm}

\noindent {\bf Acknowledgements:} It is a pleasure to thank Professor
C.R.~Doering for a very enlight\-en\-ing and stimulating discussion of the
Doering-Constantin approach. This work was supported by the Deutsche
Forschungsgemeinschaft via the Sonderforschungsbereich
``Nichtlineare Dynamik'', SFB~185, and by the German-Israeli-Foundation
(GIF).

\begin{figure}
\caption[FIG.~1] {Background flow profiles resulting from the variational
    principle with the sharpened profile constraint~(\ref{VIER_CONSTR}).
    Solid line: $0 \leq Re \leq 16\sqrt{2}$;
    short dashes: $Re = 20\sqrt{2}$,
    long dashes: $Re = 64\sqrt{2}$.}
\end{figure}

\begin{figure}
\caption[FIG.~2] {Balance parameter $a$ resulting from the variational
    principle with the sharpened profile constraint~(\ref{VIER_CONSTR}).}
\end{figure}

\begin{figure}
\caption[FIG.~3] {Bounds on the dimensionless energy dissipation rate
    $c_{\varepsilon} \! \left(Re\right)$ derived from the variational
    principle with the sharpened profile constraint~(\ref{VIER_CONSTR}).
    Solid line: lower bound, $1/Re$;
    long dashes: upper bound with optimized parameter $a$
    (see inequality~(\ref{VIER_CBOUND}));
    short dashes: upper bound obtained in ref.~\cite{GebhardtEtAl95}
    for $a = 2$ fixed (see inequality~(\ref{VIER_CBOUNDDC})).}
\end{figure}
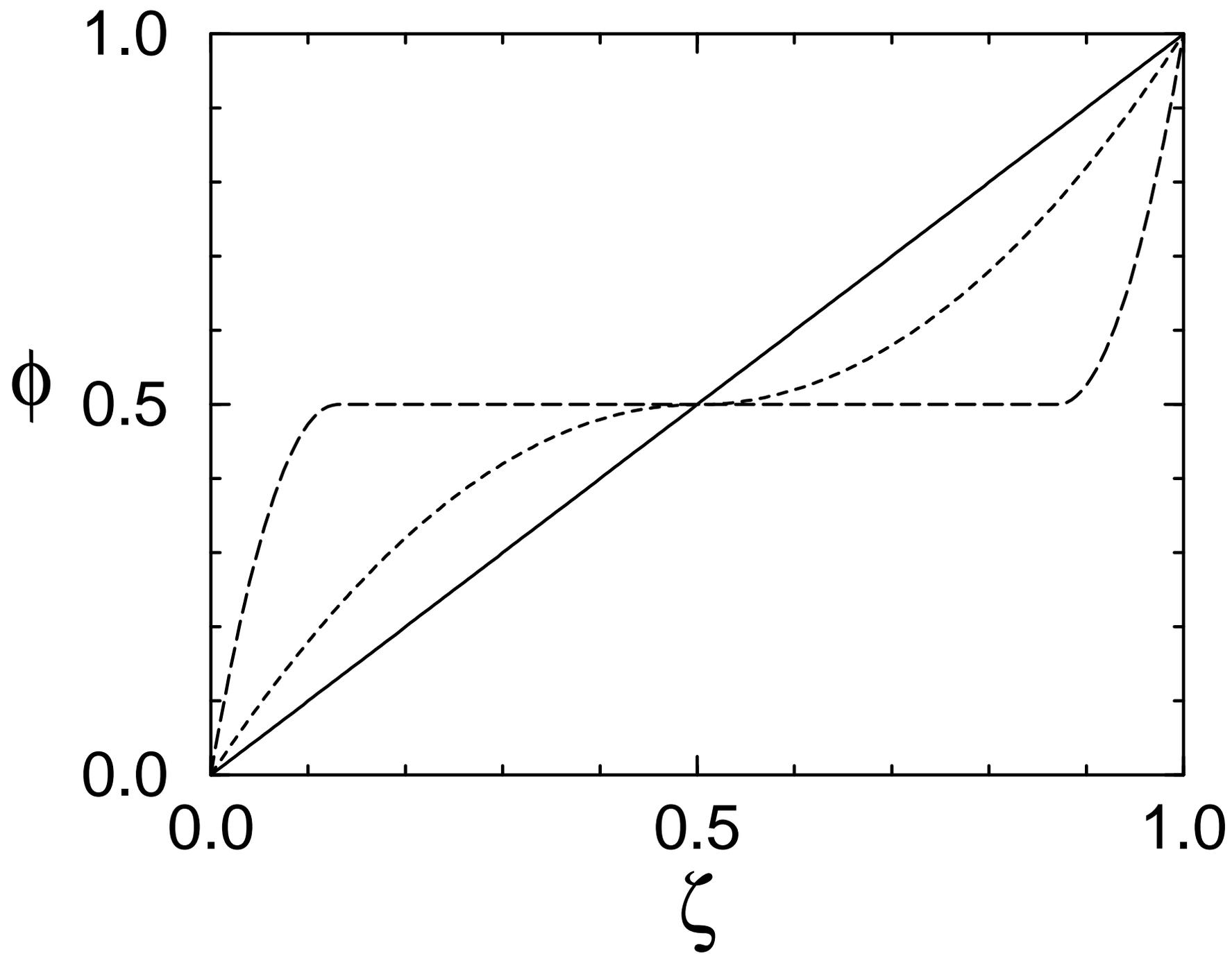
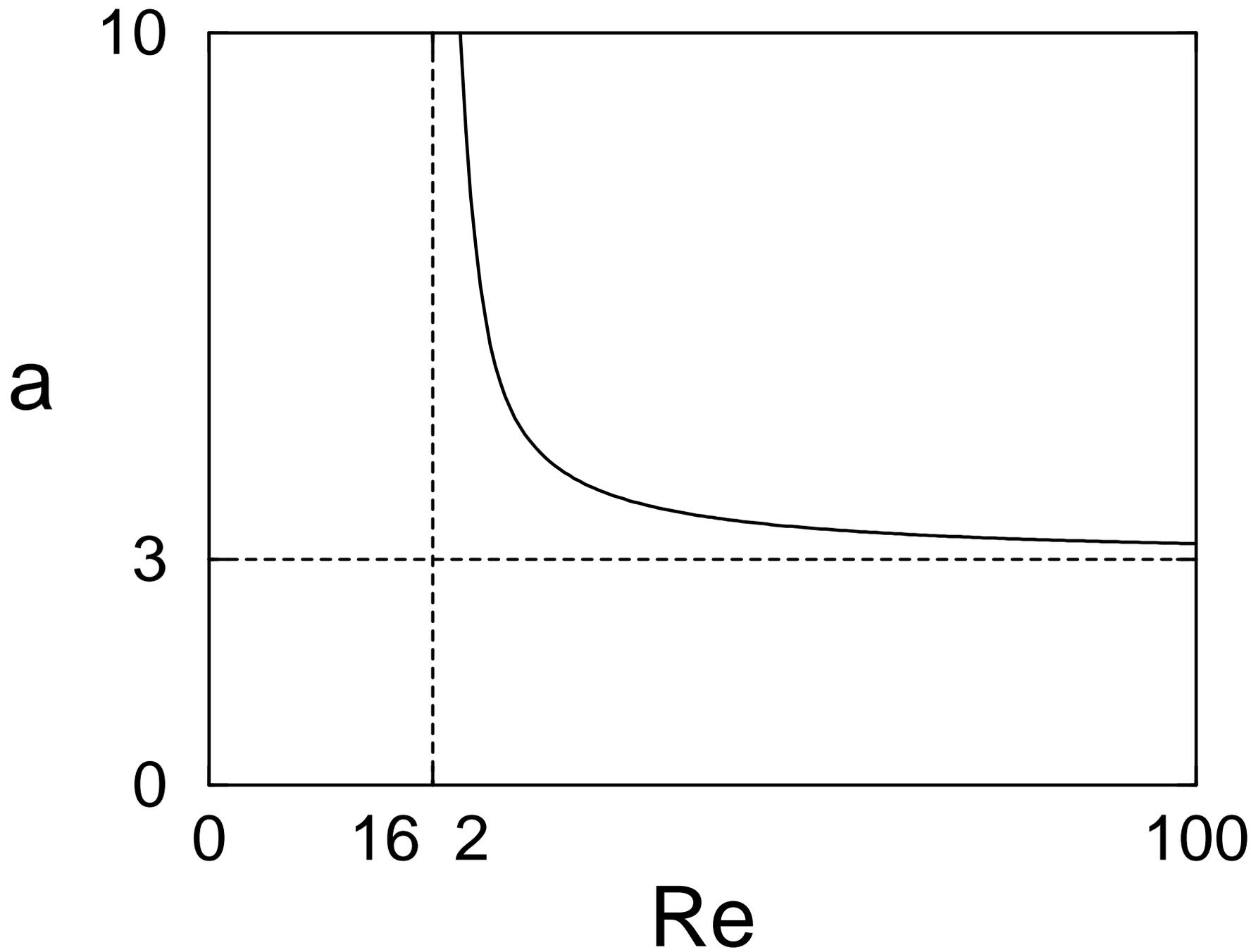
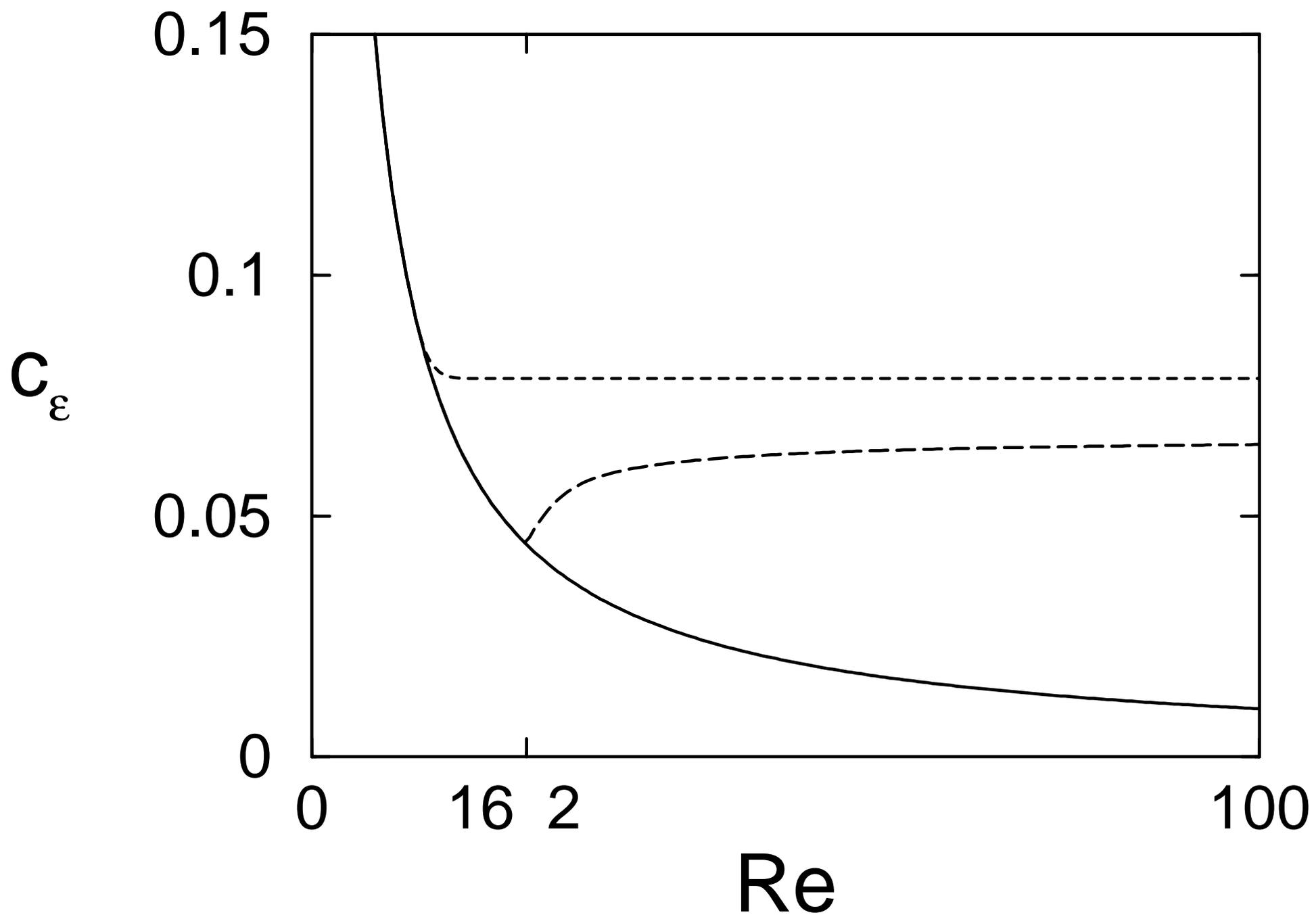
\end{document}